\documentclass[aps,pra,twocolumn,amsmath,amssymb, superscriptaddress,tightenlines]{revtex4}
\usepackage{graphicx}
\usepackage{epsfig}
\usepackage{dcolumn}
\usepackage{bm}
\usepackage{amssymb}
\usepackage{bm}
\usepackage{amsmath, bm}
\usepackage{upgreek}
\usepackage[colorlinks,citecolor=blue,linkcolor=blue,hyperindex]{hyperref}

\begin{document}

\title{Simple analog of the black-hole information paradox in quantum Hall interfaces}
\author{Ken K. W. Ma and Kun Yang}
\affiliation{National High Magnetic Field Laboratory and Department of Physics, Florida State University, Tallahassee, Florida 32306, USA}
\date{\today}


\begin{abstract}
The black hole information paradox has been hotly debated for the last few decades, without full resolution. This makes it desirable to find analogs of this paradox in simple and experimentally accessible systems, whose resolutions may shed light on this long-standing and fundamental problem. Here we identify and resolve an apparent ``information paradox" in a quantum Hall interface between the Halperin-331 and Pfaffian states. Information carried by the pseudospin degree of freedom of the Abelian 331 quasiparticles gets scrambled when they cross the interface to enter non-Abelian Pfaffian state, and becomes inaccessible to local measurements; in this sense the Pfaffian region is an analog of black hole interior while the interface plays a role similar to its horizon. We demonstrate that the ``lost" information gets recovered once the ``black hole" evaporates and the quasiparticles return to the 331 region, albeit in a highly entangled form. Such recovery is quantified by the Page curve of the entropy carried by these quasiparticles, which are analogs of Hawking radiation.

\end{abstract}

\maketitle


\section{Introduction}

The existence of black holes has received strong support from recent observations~\cite{LIGO2016, EHT2019, Paynter2021}. Instead of being a region which nothing can escape from, Hawking predicted that a black hole emits radiation and evaporates slowly~\cite{Hawking1974, Hawking1975}. He also concluded that the radiation carries no information except mass, angular momentum, and charge of the black hole~\cite{Hawking1976, soft-hair}. This result points to possible loss of information in black holes. On one hand, it is consistent with the no-hair theorem~\cite{Israel67, Israel68, Carter71}. On the other hand, quantum mechanics forbids information loss in any unitary process. This apparent contradiction leads to the black hole information paradox~\cite{Giddings, Mathur, Marolf}. It is believed that the resolution of this paradox may provide important clues on how to combine quantum mechanics and general relativity.

Various approaches have been proposed to resolve the paradox. Among them, the holographic principle~\cite{t-Hooft, Susskind, Bousso, Barbon} supports the preservation of unitarity and information. In particular, information can be encoded holographically on surfaces, such as the event horizon. This belief is substantiated by the discovery of the anti-de Sitter/conformal field theory (AdS/CFT) correspondence~\cite{Maldacena}. Recently, the firewall scenario~\cite{AMPS} was conjectured to resolve the conflict between black hole complementarity~\cite{STU, StW} and monogamy of entanglement~\cite{CKW-inequality}. If this conjecture is correct, the firewall at the event horizon (or black hole's boundary) may also break the entanglement between the outgoing and the infalling particles. Thus, the boundary can be as important as, or even more important than, the interior of a black hole.

Let us assume black hole evaporation is a unitary process. Then how is information hidden in the black hole released from Hawking radiation? Page argued that the release of information starts slowly at the beginning, but becomes faster in the later stage of the evaporation~\cite{Page1993-BH, Page1993-entropy, Page1993-review}. If the system was initially in a pure state, entropy of the radiation (coming from its entanglement with the remainder of black hole) would first increase from zero but eventually decrease back to zero when the black hole evaporates completely, thus recovering the pure state nature of the system and all the (quantum) information it carries. This feature is now known as the Page curve. Based on quantum information theory, the thought experiment by Hayden and Preskill (Hayden-Preskill protocol) has provided further insight on retrieving information from Hawking radiation~\cite{Hayden-Preskill}. Suppose the black hole has already passed its Page time and become maximally entangled with its previously emitted Hawking radiation. If the internal dynamics of black hole can be described by an instantaneous random unitary transformation, then any additional information entering the black hole can be recovered from Hawking radiation almost immediately (a very short time compared to the lifetime of the black hole)~\cite{Hayden-Preskill, Kitaev2017}. The protocol has postulated the existence of information scrambling, which has been demonstrated in recent quantum circuit experiment~\cite{Landsman2019, Yao2019}. In addition, recent studies have recovered the Page curve for AdS black holes~\cite{Penington, AEMM, PSSY, AEH, AHMST, RMP2021}. However, a full resolution of the paradox remains an open problem~\cite{Raju}. It is thus desirable to mimic the information paradox in simple and experimentally accessible systems, that allows for a complete understanding of this process.

Somewhat similar to the holographic principle, the bulk-edge correspondence relates the topologically protected edge modes and bulk topological orders in fractional quantum Hall (FQH) systems~\cite{Wen-book}. This allows us to learn about the bulk by probing the edge of the system~\cite{Dima-review2020}. Comparatively speaking, interfaces between a pair of FQH states are explored much less~\cite{Grosfeld2009, Bais-PRL2009, wan16, Yang2017, Mross, Wang, Lian, simon20, zhu20, Hughes2019,  Regnault1, Regnault2, Nielsen1, Nielsen2, Teo2020, Heiblum2021, Mross2021}. The physics of interfaces is much richer than simple edges~\cite{QH-interface2021}. For example, as we demonstrate below, certain interfaces allow quasiparticle tunneling between two different FQH states, even if the quasiparticles are of very different nature. If they have different internal degrees of freedom, (local) information carried by them needs to be transmuted (or scrambled in a specific way) to prevent information loss. This motivates us to explore analogs of black hole information paradox in quantum Hall interfaces.

In this paper, we identify and resolve an analog of information paradox in the quantum Hall interface between the Halperin-331~\cite{Halperin} and the Pfaffian (Moore-Read)~\cite{MR1991} states. The Pfaffian state is well-known since it hosts non-Abelian anyons, which may be useful in topological quantum computation~\cite{TQC-RMP2008}. Meanwhile, both 331 and Pfaffian states may describe FQH effect at half-integer filling factors. Here, we focus on the $\nu=1/2$ FQH state in bilayer systems or wide quantum wells~\cite{Suen-bilayer, Eisenstein-bilayer}. Due to the competition between interlayer tunneling and intralayer Coulomb interaction, a phase transition between the 331 and Pfaffian states was predicted~\cite{Suen1994, Papic2010, Sheng2016}. This suggests the possibility of creating a 331-Pfaffian interface by controlling the tunneling strengths in different regions of the bilayer system. Interestingly, the 331 state has a pseudospin (layer) degree of freedom, which is absent in the Pfaffian state. If the original information carried by the pseudospin degree of freedom becomes irrecoverable after quasiparticles cross the interface and enter the Pfaffian liquid, it leads to an ``information paradox". We demonstrate that the information is scrambled and stored nonlocally in the Pfaffian liquid and the interface. We also mimic black hole evaporation in the same system, and find it satisfies the Page curve naturally. In other words, the original pseudospin information is recovered and the ``information paradox" in our model is resolved. Here, we need to emphasize that we are not aiming at a resolution of the original information paradox in astrophysical black holes. This is clearly unachievable by proposing a simple analogy. Instead, we want to simulate some important concepts in resolving the original paradox in a simple and accessible manner.

\section{The 331-Pfaffian interface and information paradox}
\label{sec:331-Pf}

\subsection{A brief review of quantum Hall effect}

To set the stage for later discussion, we first review briefly some basic concepts in quantum Hall (QH) physics~\cite{Yang-book}. Electrons moving in two dimensions ($x-y$ plane) and a perpendicular magnetic field (in the $z$ direction) have their energy levels being quantized in Landau levels. Depending on the ratio between the number of electrons and the number of magnetic flux quanta enclosed by the system, the system can have different filling factors
$\nu$. Of particular interest is the case $\nu<1$, where only the lowest Landau level is partially filled by electrons at low temperature. Since the kinetic energy of these electrons is quenched due to Landau quantization, the interaction between them dominates the properties of the system. Various FQH states, which possess numerous fascinating properties, are realized in this strongly correlated electronic system. The exotic properties of FQH states are associated with the topological order they possess~\cite{Wen-book}. Most prominent among them is the existence of low-energy excitations (quasiparticles) that have fractional charges and obey fractional statistics (between bosonic and fermionic statistics)~\cite{review-fractional}. A famous example is the Laughlin state at $\nu=1/3$~\cite{Tsui1982}, in which quasiparticle with a fractional charge $e/3$ and a fractional statistics $2\pi/3$ can exist~\cite{Laughlin, Arovas}. Note that both fractional charge and fractional statistics were observed experimentally~\cite{de-Picciotto, Saminadayar, Bartolomei}. Such exotic quasiparticles are called anyons, and the possible types of anyons are associated with the specific topological order.

The bulk-edge correspondence, another consequence of the topological order, relates the edge structure and the bulk topological order in FQH systems. In particular, it predicts the existence of gapless edge modes described by conformal field theories (CFTs), and there is a one-to-one correspondences between the bulk topological order and edge CFT~\cite{CFT-QHE}. In our previous example, the edge of the Laughlin state at $\nu=1/3$ has a single chiral bosonic edge mode $\phi$, which can be described by the Lagrangian density,
\begin{eqnarray}
\mathcal{L}_{1/3}
=-\frac{3}{4\pi}\partial_x\phi(\partial_t-v\partial_x)\phi.
\end{eqnarray}
Here, $v$ is the speed of the edge mode, and $\phi$ is a (chiral) bosonic field. In general, the edge of a FQH liquid can have more than one edge mode. For Abelian FQH states, the corresponding edge theory is described by~\cite{Wen-book}
\begin{eqnarray}
\mathcal{L}_{\rm edge}
=-\frac{1}{4\pi}\sum_{i,j}K_{ij}\partial_t\phi_i\partial_x\phi_j
-\frac{1}{4\pi}\sum_{i,j}V_{ij}\partial_x\phi_i\partial_x\phi_j.
\end{eqnarray}
Importantly, the $K$ matrix encodes all information of the topological order. For a FQH state in a bilayer system, it may (but not always) be described by a two-component topological order which has a $2\times 2$ $K$ matrix. In this situation, two different edge modes exist. Furthermore, the possible type of edge modes is not limited to bosonic mode. Other types of modes such as Majorana fermion modes exist when the topological orders are non-Abelian~\cite{MR-edge, Levin-APf, Lee-APf, ZF2016}.

With the knowledge of the edge structure in hand, different low-energy excitations in the FQH system can be described or created by suitable CFT operators~\cite{CFT-QHE}. For example, a charge-$e/3$ quasiparticle and an electron in the $\nu=1/3$ Laughlin state are created by the operators $:\exp{(i\phi}):$ and $:\exp{(3i\phi}):$, respectively. Here, $:\mathcal{V}:$ denotes the normal ordering of the vertex operator $\mathcal{V}$. When there is no confusion, this normal ordering notation will be dropped in the later discussion. For a FQH state being described by a multicomponent topological order, there are multiple types of anyons (described by different CFT operators) that have the same electric charge. In other words, the anyons have an additional degree of freedom. This point will become clear when we discuss our setup.

\subsection{The 331-Pfaffian interface as a firewall}

\begin{figure} [htb]
\includegraphics[width=3.3in]{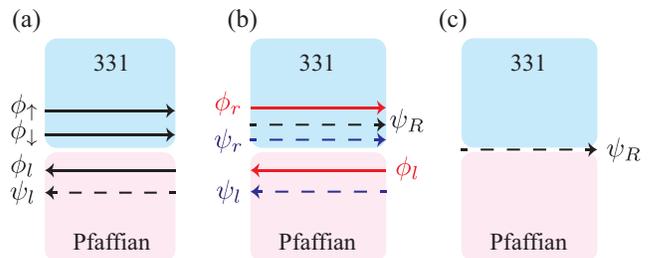}
\caption{Localization of edge modes at the interface due to random electron tunneling between 331 edge and Pfaffian edge. Solid lines denote charge modes, whereas dashed lines correspond to neutral modes. (a) The original edge modes in the 331 liquid and Pfaffian liquid. (b) Counterpropagating edge modes with the same color are gapped or localized. (c) Only a single chiral Majorana fermion mode remains gapless and propagating along the interface.}
\label{fig:interface}
\end{figure}

The specific system we consider is the interface between Halperin-331 and Pfaffian quantum Hall liquids. Both QH liquids have the same Landau level filling factor $\nu=1/2$, which can be realized in a bilayer system. For the 331 liquid, it is described by a two-component topological order with the $K$ matrix~\cite{Halperin, Wen-book},
\begin{eqnarray}
K=\begin{pmatrix}
3 & 1 \\
1 & 3
\end{pmatrix}.
\end{eqnarray}
The two different edge modes are denoted as $\phi_\uparrow$ and $\phi_\downarrow$. The two most relevant operators creating an electron are $\exp{(3i\phi_\uparrow+i\phi_\downarrow)}$ and $\exp{(i\phi_\uparrow+3i\phi_\downarrow)}$. On the other hand, the Pfaffian liquid is described by a single-component non-Abelian order with $K=2$. Its edge has a bosonic mode $\phi$ and a Majorana fermion mode $\psi$~\cite{MR-edge}. The corresponding electron operator is $\psi\exp{(2i\phi)}$. Since the Halperin-331 and Pfaffian edges have opposite chiralities at the interface, the interface is described by the Lagrangian density~\cite{Yang2017},
\begin{align} \label{eq:331-Pf}
\nonumber
\mathcal{L}
=&-\frac{1}{4\pi}\sum_{i,j}K_{ij}\partial_t\phi_i\partial_x\phi_j
+\frac{2}{4\pi}\partial_t\phi_l\partial_x\phi_l
-i\psi_l\partial_t\psi_l
\\
&-\mathcal{H}(\phi, \psi).
\end{align}
Here, the indices $i,j=\uparrow, \downarrow$ denote the layer or the pseudospin.
All $\phi_\uparrow$, $\phi_\downarrow$, and $\phi_l$ are charge modes. The first two are right-moving along the edge of the 331 liquid, whereas the last one is left-moving along the edge of the Pfaffian liquid. The Pfaffian liquid also has a left-moving neutral Majorana fermion mode $\psi_l$ along the edge. The edge structures of both quantum Hall states are illustrated in Fig.~\ref{fig:interface}(a). As shown by one of us, a relevant random electron tunneling between the Pfaffian and 331 edges can lead to a phase transition at the interface~\cite{Yang2017}.

Now, we follow Ref.~\cite{Yang2017} and briefly summarize how different modes get localized at the interface. This also allows us to introduce useful notations for later discussion. One can define a charge mode $\phi_r=\phi_\uparrow+\phi_\downarrow$ and a neutral spin mode $\phi_n=\phi_\uparrow-\phi_\downarrow$ in the 331 liquid. Using this new set of modes, the topological term of the 331 edge becomes
\begin{eqnarray}
\mathcal{L}_{331}
=-\frac{2}{4\pi}\partial_t\phi_r\partial_x\phi_r
-\frac{1}{4\pi}\partial_t\phi_n\partial_x\phi_n.
\end{eqnarray}
The overall charge density at the interface is given by
\begin{eqnarray}
\rho(x)
=\frac{1}{2\pi}\partial_x(\phi_\uparrow+\phi_\downarrow+\phi_l)
=\frac{1}{2\pi}\partial_x\phi_c.
\end{eqnarray}
The random electron tunneling between the Pfaffian and 331 edges is described by
\begin{align} \label{eq:mode-gap}
\nonumber
H_{\rm T}
=&\int \xi(x)\psi_l
\left(e^{3i\phi_\uparrow+i\phi_\downarrow+2i\phi_l}
+e^{i\phi_\uparrow+3i\phi_\downarrow+2i\phi_l}\right)~dx
+\text{H.c.}
\\
=&\int |\xi(x)|\psi_l(x) \psi_r(x)\cos{\left[2\phi_c(x)+\varphi(x)\right]}~dx.
\end{align}
Here, $\xi(x)$ denotes the random tunneling amplitude. In the second line, $|\xi(x)|$ and
$\varphi(x)$ are the magnitude and the phase of $\xi(x)$, respectively. We have also fermionized $\exp{[i\phi_n(x)]}=\psi_r(x)+i\psi_R(x)$. The resulting edge modes are shown in Fig.~\ref{fig:interface}(b). If $H_{\rm T}$ is relevant in the renormalization group sense, then both charge modes, $\psi_l$, and $\psi_r$ are localized at the interface. After the localization, only a single right-moving Majorana fermion mode remains gapless and propagates freely at long distance, as shown in Fig.~\ref{fig:interface}(c). Notice that this gapless mode is neither an original edge mode of the 331 nor the Pfaffian state.

Since Pfaffian and 331 states are quantum Hall states formed by superconducting pairing between composite fermions~\cite{Read-Green}, both of these two states have quasiparticles with the smallest possible charge of $e/4$~\cite{MR1991, Nayak-Wilczek}. However, there is a fundamental difference between these quasiparticles. For the 331 state, there are two different types of Abelian $e/4$ quasiparticles created by the vertex operators, $e^{i\phi_\uparrow}$ and $e^{i\phi_\downarrow}$. One may view them as quasiparticles with different pseudospins. When we formulate the information paradox in the following discussion, this pseudospin will be regarded as the \textit{degree of freedom} of the Abelian quasiparticles. For the Pfaffian state, there is only one type of $e/4$ quasiparticle created by $\sigma e^{-i\phi_l/2}$~\cite{MR1991}. Here, $\sigma$ is the twist field with a scaling dimension $1/16$ in the chiral Ising CFT~\cite{CFT-book}. We summarize the three primary fields and their properties in Table~\ref{tab:CFT-Ising}. In particular, $\sigma$ satisfies the fusion rule $\sigma\times\sigma=\psi+I$. Note that we have omitted the subscript $l$ for the Majorana field to make the discussion of Ising CFT general. Its proper meaning should be clear from context. The fusion rule indicates that the quasiparticle is non-Abelian. An interesting question is what happens if a quasiparticle is dragged from the 331 liquid in to the Pfaffian liquid? It seems that the pseudospin information would be lost. In this sense, we can define the interface between the two different QH liquids as the \textit{``event horizon with a firewall"} in our setup. This definition or analogy makes sense since the interface plays the role of a one-way surface of information in our setup, and the ``destruction" of pseudospin information at the interface resembles a firewall conjectured in Ref.~\cite{AMPS}. In addition, the Pfaffian liquid can be viewed as the interior of a \textit{``black hole"}. Suppose this analogous black hole can evaporate (discussed in Sec.~\ref{sec:evaporation}) and the original pseudospin information cannot be recovered at the end of the evaporation. Then, the lost of information contradicts to the fact that quasiparticle tunneling is a unitary process. Thus, we have identified an apparent ``information paradox".

\begin{table} [htb]
\renewcommand{\arraystretch}{1.2}
\begin{tabular}{|c|c|c|}
\hline
~Primary field~ & ~Conformal spin~ & ~Quantum dimension~
\\ \hline
~$I$~ & $0$ & ~$1$~
\\
~$\psi$~   & $1/2$ & ~$1$~
\\
~$\sigma$~  & $1/16$ & ~$\sqrt{2}$~
\\ \hline
\end{tabular}
\caption{Primary fields in the chiral Ising CFT with a central charge $c=1/2$.}
\label{tab:CFT-Ising}
\end{table}

\section{331-Pfaffian interface from anyon condensation}
\label{sec:interface-condensation}

Before resolving the paradox, we reformulate the above discussion in the framework of anyon condensation~\cite{Bais-PRB2009, Bais-PRL2009, Ellens2014, Burnell-review, Bernevig}. This technique has been commonly applied to study possible transitions between topologically ordered phases. In the context of quantum Hall physics, it was used to study the interface between Pfaffian and non-Abelian spin-singlet (NASS) quantum Hall states~\cite{Bais-PRL2009, Grosfeld2009}. In this section, we first use anyon condensation to deduce the CFT description of the 331-Pfaffian interface. In the next section, we apply the same technique to resolve the paradox. Along the way, we adopt a pedagogical approach and aim at relating the rather abstract technique to the more physical picture in Sec.~\ref{sec:331-Pf}. It will allow us to highlight the advantages of applying anyon condensation in studying quantum Hall interfaces.

From Sec.~\ref{sec:331-Pf}, we know that the edges of the Halperin-331 and Pfaffian liquids are described by CFTs with central charges $2$ and $3/2$, respectively. These two edges are \textit{counterpropagating} at the interface. Hence, we expect the resulting CFT describing the 331-Pfaffian interface has a net central charge of $2-3/2=1/2$. To deduce exactly what the CFT is, it is first necessary to separate the charge and neutral sectors for both Halperin-331 and Pfaffian liquids. It is because a charge mode cannot be gapped out by coupling to a neutral mode in a usual situation. Equivalently, we do not consider the possibility of condensing charge bosons, which will break the U(1) gauge symmetry. The separation was already achieved in Sec.~\ref{sec:331-Pf}. In particular, the combination of charge modes $\phi_c=\phi_r+\phi_l$ was shown to be gapped out (more precisely, localized) by $H_{\rm T}$~\cite{footnote-gap}. Therefore, we can focus our discussion on the neutral sectors.

As stated previously, the neutral sector of the Pfaffian state is described by a chiral Ising CFT. For the Halperin-331 state, its neutral sector is governed by the spin mode $\phi_n$, which is described by the U(1)$_4$ CFT. Different primary fields in this Abelian CFT are summarized in Table~\ref{tab:CFT-vertex}. Note that any two vertex operators in the form
$e^{i\alpha\phi_n/2}$ and $e^{i(\alpha+4\mathbb{Z})\phi_n/2}$ are identified.

\begin{table} [htb]
\renewcommand{\arraystretch}{1.2}
\begin{tabular}{|c|c|c|c|}
\hline
~Symbol~ & ~Vertex operator~ & ~Conformal spin~ & ~Type~
\\ \hline
$\mathcal{V}_0$  & $1$ & $0$ & ~~Boson~~
\\
$\mathcal{V}_1$  & $\exp{(i\phi_n/2)}$ & $1/8$ & ~~Anyon~~
\\
$\mathcal{V}_2$  & $\exp{(i\phi_n)}$ & $1/2$ & ~~Fermion~~
\\
$\mathcal{V}_3$  & $\exp{(3i\phi_n/2)}$ & $1/8$ & ~~Anyon~~
\\ \hline
\end{tabular}
\caption{Primary fields in the U(1)$_4$ CFT. Here, the normal ordering in the vertex operators are not shown explicitly. Note that $\mathcal{V}_3=\exp{(3i\phi_n/2)}\simeq \exp{(-i\phi_n/2)}$.}
\label{tab:CFT-vertex}
\end{table}

The structure (remaining gapless modes) of the 331-Pfaffian interface is solely determined by anyon condensation in the neutral sectors. This condensation occurs in the
$\text{U(1)}_4 \times \overline{\text{Ising}}$ CFT. We emphasize again that the bar denotes \textit{conjugation} of the Ising CFT due to the opposite chiralities between the 331 and Pfaffian edges at the interface.  Compared to the original CFT, anyons in the conjugate CFT have the same fusion rules, but complex conjugated topological spins and braiding phases. Alternatively, one may interpret the condensation as a coset construction~\cite{Bais-PRB2009}. We label a generic anyon as $(e^{i\alpha\phi_n/2}, \bar{t})$. Here, the parameter
$\alpha=0,1,2,3$ determines the corresponding primary fields in the U(1)$_4$ CFT. Meanwhile, $\bar{t}=\left\{\bar{I}, \bar{\psi}, \bar{\sigma}\right\}$ denotes the primary fields in the conjugate Ising CFT. In the present case, there is only one condensable boson,
$B=(e^{i\phi_n}, \bar{\psi})$. The condensation of $B$ leads to confinement of some of the anyons in the condensed phase. An anyon remains unconfined if and only if it has a trivial mutual statistics with $B$. This condition ensures that an unconfined anyon has a consistent topological spin in the condensed phase. Furthermore, two anyons are identified when they differ from each other by a multiple of $B$. Using operator product expansion, it is straightforward to deduce the six (or three after identification) deconfined anyons in the condensed phase. They are listed in Table~\ref{tab:confined}. Their corresponding topological sectors, namely $\tilde{I}$, $\tilde{\psi}$, and $\tilde{\sigma}$ are defined according to their conformal spins. From the table, we conclude that the 331-Pfaffian interface is described by a chiral Ising CFT. Note that this Ising CFT has an opposite chirality to the one describing the Pfaffian edge at the interface.

\begin{table} [htb]
\renewcommand{\arraystretch}{1.4}
\begin{tabular}{|c|c|c|}
\hline
~Sector~ & ~Unconfined anyons~ & ~Conformal spin~
\\ \hline
$\tilde{I}$  & ~~$(\mathcal{V}_0, \bar{I})\simeq(\mathcal{V}_2, \bar{\psi})$~~ & $0$
\\
$\tilde{\psi}$  & ~~$(\mathcal{V}_0, \bar{\psi})\simeq(\mathcal{V}_2, \bar{I})$~~ & $1/2$
\\
$\tilde{\sigma}$  & ~~$(\mathcal{V}_1, \bar{\sigma})\simeq(\mathcal{V}_3, \bar{\sigma})$~~ & $1/16$
\\ \hline
\end{tabular}
\caption{Unconfined anyons in the $\text{U(1)}_4\times\overline{\text{Ising}}$ CFT after condensing the boson $B=(e^{i\phi_n}, \bar{\psi})$. Vertex operators $\mathcal{V}_i$ are defined in Table~\ref{tab:CFT-vertex}. Here, the symbol $\simeq$ denotes identification of anyons modulo $B$. The conformal spins are deduced from $s=h_1-h_2~(\text{mod}~1)$, where $h_1$ and $h_2$ denote the conformal dimensions of primary fields in the U(1)$_4$ and Ising CFTs, respectively.}
\label{tab:confined}
\end{table}

Now, one may wonder why going through such abstract and seemingly redundant procedures to find out the CFT describing the interface. Doesn't the net central charge $c=1/2$ directly indicate that it should be an Ising CFT? There are two reasons for analyzing this simple system by anyon condensation. First of all, it is fortunate that for the 331-Pfaffian interface, the mechanism and consequences of anyon condensation can be visualized in a very transparent and physical manner, but this is a very special case. In Eq.~\eqref{eq:mode-gap}, the electron tunneling between counterpropagating edges at the interface couples $\psi_l$ and $e^{i\phi_n}$. This leads to a mass term and eventually gaps out the counterpropagating $\psi_l$ and $\psi_r$. Only $\psi_R$ remains gapless at the interface. This was demonstrated by fermionizing $e^{i\phi_n}=\psi_r+i\psi_R$. This type of arguments does not always generalize to more complicated interfaces. On the other hand, the condensation of $B$ systematically captures the gaping process and leads to a correct CFT description of the 331-Pfaffian interface. More importantly, anyon condensation relates every primary field in the original and condensed phases. These relations cannot be obtained from the argument in Sec.~\ref{sec:331-Pf}.

\section{Transmutation of pseudospin information} \label{sec:resolution}

The previous section has set the stage for us to discuss the transmutation of pseudospin information when Abelian quasiparticles cross the interface.

We first discuss and comment on the charge sectors. As one will see, they basically play no role in the resolution of the paradox. Since charge-$e/4$ quasiparticles are allowed in both Halperin-331 and Pfaffian liquids, dragging quasiparticles across the interface does not require any absorption of net charge at the interface. Furthermore, the gapping of $\phi_c$ indicates that the dragging will not create any low-energy charge excitation at the interface~\cite{footnote0}.

We thus focus on the neutral sectors. An Abelian quasiparticle in the Halperin-331 liquid has its pseudospin degree of freedom carried solely by the neutral mode $\phi_n$. This is observed by writing
\begin{align}
e^{i\phi_\uparrow}
&=e^{i\phi_r/2}e^{i\phi_n/2},
\\
e^{i\phi_\downarrow}
&=e^{i\phi_r/2}e^{-i\phi_n/2}.
\end{align}
Hence, the vertex operators $\mathcal{V}_1$ and $\mathcal{V}_3$ encode the spin-up and spin-down states of an Abelian charge-$e/4$ quasiparticle, respectively. These two operators are not defined in the Pfaffian liquid. To understand the transmutation of quasiparticles when they cross the interface, we need to represent the four primary fields in the U(1)$_4$ CFT as different products between two Ising CFTs. One of them describes the interface, whereas the other describes the Pfaffian order. Both Ising CFTs now have the same chirality to match the central charges, $1=1/2+1/2$. From Table~\ref{tab:confined}, one can obtain the inverted expressions:
\begin{align}
\label{eq:inverted}
\mathcal{V}_0
&\equiv I_1
=I_{1/2}\times \tilde{I}+\psi\times \tilde{\psi},
\\
\label{eq:inverted1}
\mathcal{V}_1
&\equiv e^{i\phi_n/2}
=\sigma\times\tilde{\sigma},
\\
\label{eq:inverted-2}
\mathcal{V}_2
&\equiv e^{i\phi_n}
=\psi\times\tilde{I}+I_{1/2}\times\tilde{\psi},
\\
\label{eq:inverted3}
\mathcal{V}_3
&\equiv e^{-i\phi_n/2}
=\sigma\times\tilde{\sigma}.
\end{align}
The tilded and untilded fields are in the CFTs describing the interface and the Pfaffian liquid, respectively. Also, the subscripts $1$ and $1/2$ in the identity fields denote the central charges of the corresponding CFTs. When there is no confusion, these subscripts will be skipped.

\begin{figure} [htb]
\includegraphics[width=3.3in]{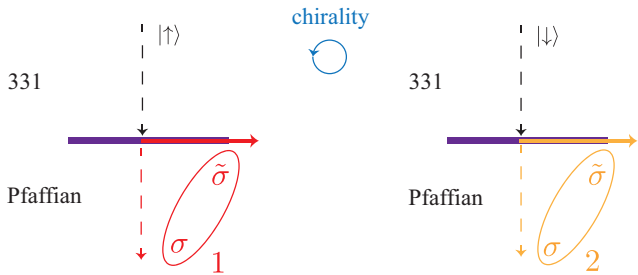}
\caption{Transmutation of an Abelian 331 quasiparticle when it crosses the interface and enters the Pfaffian liquid. Here, only the neutral sector is considered (see the main text for more details). The symbols $|\uparrow\rangle$ and $|\downarrow\rangle$ denote quasiparticles with pseudospin up and down, respectively. Their corresponding vertex operators are $\mathcal{V}_1$ and $\mathcal{V}_3$ in the U(1)$_4$ CFT.}
\label{fig:pseudospin}
\end{figure}

Eq.~\eqref{eq:inverted} suggests that the U(1)$_4$ CFT is obtained from condensing the boson $b=(\psi,\tilde{\psi})$~\cite{Ardonne-spin, Ardonne-para}. This result is consistent with the orbifold construction~\cite{DVVV1989}. After the condensation, one of the unconfined particles is $(\sigma, \tilde{\sigma})$. We should state clearly that these two twist fields describe excitations (anyons) at \textit{different} regions of the system, so it is meaningless to consider their fusion. In other words, the present situation is different from the case of a Pfaffian liquid, in which $\sigma$ in the bulk and $\sigma$ at the edge created from vacuum must fuse to $I$ for conserving fermion parity. Importantly,
\begin{eqnarray} \label{eq:sigma-fuse}
(\sigma, \tilde{\sigma})\times(\sigma, \tilde{\sigma})
=(\psi, \tilde{\psi})+(I, \tilde{I})+(\psi, \tilde{I})+(I, \tilde{\psi}).
\end{eqnarray}
The first two terms on the right hand side show that two orthogonal copies of vacuum exist, so $(\sigma, \tilde{\sigma})$ needs to split into two \textit{inequivalent} types of anyons in the resulting U(1)$_4$ CFT~\cite{Ardonne-spin}. We denote them as $(\sigma, \tilde{\sigma})_1$ and $(\sigma, \tilde{\sigma})_2$. Both of them have conformal spins $1/8$, which are identified as the vertex operators $\mathcal{V}_1$ and $\mathcal{V}_3$ in the U(1)$_4$ CFT  (see Table~\ref{tab:CFT-vertex}). The fusion rules are consistent by imposing the conditions
$(\sigma, \tilde{\sigma})_1\times(\sigma, \tilde{\sigma})_1=
(\sigma, \tilde{\sigma})_2\times(\sigma, \tilde{\sigma})_2
=\mathcal{V}_2$
and $(\sigma, \tilde{\sigma})_1\times(\sigma, \tilde{\sigma})_2=\mathcal{V}_0$. Following Ref.~\cite{Bais-PRL2009}, we interpret the above result as an incoming pseudospin-up quasiparticle transmutes into a neutral anyon $\tilde{\sigma}$ at the interface, and another anyon $\sigma$ free to move in the Pfaffian liquid. To be more precise, the last anyon actually carries charge $e/4$, but we skip displaying its charge sector $e^{-i\phi_l/2}$ explicitly. The same conclusion holds for an incoming quasiparticle with pseudospin down. We illustrate the results in Fig.~\ref{fig:pseudospin}.

\subsection{Matching of Hilbert spaces and analogy of information scrambling}
\label{sec:matching}

It is obvious that the total quantum dimension of $(\sigma, \tilde{\sigma})_1$ and $(\sigma, \tilde{\sigma})_2$ is two. It matches the two-dimensional Hilbert space spanned by the pseudospin degree of freedom of an Abelian charge-$e/4$ quasiparticle. This matching is guaranteed mathematically by the commutativity between fusion and restriction in anyon condensation~\cite{Bais-PRB2009}. Interestingly, the information of pseudospin is being stored \textit{nonlocally} at the interface \textit{and} in the interior of Pfaffian liquid. There is no local measurement that can distinguish between $(\sigma, \tilde{\sigma})_1$ and $(\sigma, \tilde{\sigma})_2$. Hence, it is impossible to recover the original information from any local measurement. This feature resembles a quantum information scrambling, which can be defined as the spreading of local information into many-body entanglement and correlation in the whole system~\cite{Landsman2019}.

The situation becomes more interesting when we keep dragging more Abelian quasiparticles across the interface. Suppose $N-1$ charge-$e/4$ quasiparticles were already dragged. We assume the corresponding $N-1$ anyons $\tilde{\sigma}$ created at the interface are well separated from each other, so that no fusion occurs between them. We also pose the same assumption for the $N-1$ anyons created in the Pfaffian liquid. Consider dragging an additional charge-$e/4$ quasiparticle across the interface. This process increases both numbers of $\tilde{\sigma}$ and $\sigma$ by one. As a result, there are $N$ neutral anyons
$\tilde{\sigma}$ at the interface, and $N$ non-Abelian anyons in the interior of the Pfaffian liquid. The dimension of the corresponding topological Hilbert space is then increased by a factor of two, which is consistent with the one bit of information carried by the additional Abelian quasiparticle from the Halperin-331 liquid. We illustrate the example of $N=6$ in Figs.~\ref{fig:evaporation}(a) and~\ref{fig:evaporation}(b). Now, we relax the confining potential, and allow the anyons to move and braid~\cite{footnote}. The braiding can further scramble the original information~\cite{Chamon2019}. The $N$ anyons at the interface are indistinguishable, so are those $N$ anyons in the Pfaffian liquid. Meanwhile, the information carried by pseudospins of the original $N$ Abelian quasiparticles is still preserved. Both Hilbert spaces for indistinguishable anyons at the interface and indistinguishable anyons in the Pfaffian liquid have dimensions $2^{N/2}$. It is intriguing that the interface and the Pfaffian liquid store the same amount of information. This does not hold in the Pfaffian-NASS interface~\cite{Bais-PRL2009, Grosfeld2009, footnote2}.

The above discussion also suggests another important feature. In addition to being stored nonlocally, the original pseudospin information is actually ``hidden" in the fusion channels of the non-Abelian anyons. Hence, the scrambled information is protected topologically and will not be destroyed by any local perturbation. This property is essential in topological quantum computation (TQC)~\cite{Preskill-notes, TQC, Kitaev2003, TQC-RMP2008, foot-TQC}.

\subsection{Upper bound of information storage and holographic principle}
\label{sec:bound}

Our previous discussion assumed that local anyons in the system can be well separated to prevent fusion. This assumption leads to a natural question. How much information can be stored nonlocally with the topological protection that has been described?

Recall that the minimum separation between two anyons is in the order of the magnetic length $\ell_B$, so that they are well defined individually and do not fuse. From this, one may naively think that the maximum amount of information can be stored is $N_A\sim A/\pi \ell_B^2$, where $A$ denotes the area of the Pfaffian liquid. This argument is valid if the information is carried solely by anyons in the Pfaffian liquid. However, this is not true in the present case. We have assumed both Pfaffian liquid and 331-Pfaffian interface were initially in the ground state with no excitations. As we discussed, the nonlocal storage of pseudospin information of the Abelian quasiparticles from the Halperin-331 liquid requires both anyons at the interface and in the Pfaffian liquid.

For an interface with a length (perimeter) $L$, it can only accommodate $N_L\sim L/\ell_B$ neutral anyons $\tilde{\sigma}$. Since the radius $R$ of a circular quantum Hall droplet satisfies $R\gg \ell_B$, one has $N_L\ll N_A$. When the number of $\tilde{\sigma}$ gets close to or exceeds $N_L$, different $\tilde{\sigma}$ anyons start to fuse. The resulting particles will be either a fermion or a boson that can propagate back to the Halperin-331 liquid. More explicitly, one has
\begin{eqnarray}
(I,\tilde{\sigma})\times(I,\tilde{\sigma})
=(I,\tilde{I})+(I,\tilde{\psi}).
\end{eqnarray}
For the first fusion outcome, the two $\tilde{\sigma}$ anyons at the interface can fuse to a neutral boson with its spin part described by $\mathcal{V}_0$ [see Eq.~\eqref{eq:inverted}]. This neutral boson can then split to a pair of quasihole and quasiparticle with opposite charges but the same pseudospin, and propagate in the 331 liquid. For the second fusion outcome, the neutral fermion can split to a pair of quasihole and quasiparticle with opposite pseudospins propagating in the 331 liquid. Consequently, some of the hidden information is released and being accessible by local probes. Therefore, the ``black hole'' is no longer completely black. Note that the released information is not protected topologically and can suffer from quantum decoherence. The discussion shows that the length of the interface sets an upper bound of storing information nonlocally and topologically via $(\sigma,\tilde{\sigma})$ pairs. Furthermore, the magnetic length $\ell_B=\sqrt{1/eB}$ (in the unit of $\hbar=c=1)$ plays the role of Planck length in the present system. Here, $B$ denotes the magnetic field.

The above observation actually resembles the argument from holographic principle in black holes. Based on this principle, the maximum amount of information can be stored in a black hole is not determined by its volume, but bounded by its area~\cite{Bousso}. This is because the Bekenstein entropy of the black hole is proportional to its area~\cite{Hawking1974, Bekenstein72, Bekenstein73, Bekenstein74}, which limits the number of degree of freedoms the black hole can have. In contrast to the quantum Hall interface, the black hole can always store and ``hide" more information by increasing its area. Since the length of the quantum Hall interface is assumed to be fixed, the analogous (a weaker version of) holographic principle there implies that Hawking radiation in the form of Abelian quasiparticles and quasiholes will be released when the bound $N_L\sim L/\ell_B$ is reached. Roughly speaking, any additional incoming information is thus reflected by the interface (event horizon).

\subsection{Firewall to electrons}

So far, we have only focused on the charge-$e/4$ quasiparticles. Some readers may argue that these anyons are not fundamental particles, so their transmutation at the interface is not that unusual. In fact, the interface can also cause a dramatic effect to incoming electrons from the Halperin-331 liquid. Since $e^{3i\phi_\uparrow+i\phi_\downarrow}=e^{2i\phi_r}e^{i\phi_n}$ and $e^{i\phi_\uparrow+3i\phi_\downarrow}=e^{2i\phi_r}e^{-i\phi_n}$, the vertex operator $\mathcal{V}_2$ encodes the fermionic nature of the electron. Similar to the usual case in which the Pfaffian and the Halperin-331 liquids are separated by the vacuum, the electron can tunnel in to the Pfaffian liquid as a fermion (with a topological sector $\psi$). However, Eq.~\eqref{eq:inverted-2} implies that it is also possible for the electron to excite a Majorana fermion $\tilde{\psi}$ at the interface, leave its Fermi statistics there, and tunnel in to the Pfaffian liquid as a boson (with a topological sector $I$). In this nontrivial case, the electron cannot simply pass through the 331-Pfaffian interface as if there is nothing there. This feature is another manifestation of the firewall nature of the interface. We should note that a similar fractionalization of electron was also proposed in the interface between a $\mathbb{Z}_2$ short-ranged resonating bond quantum spin liquid and a superconductor~\cite{BBK2014}.

\section{Simulation of black hole evaporation} \label{sec:evaporation}

\begin{figure} [htb]
\includegraphics[width=3.3in]{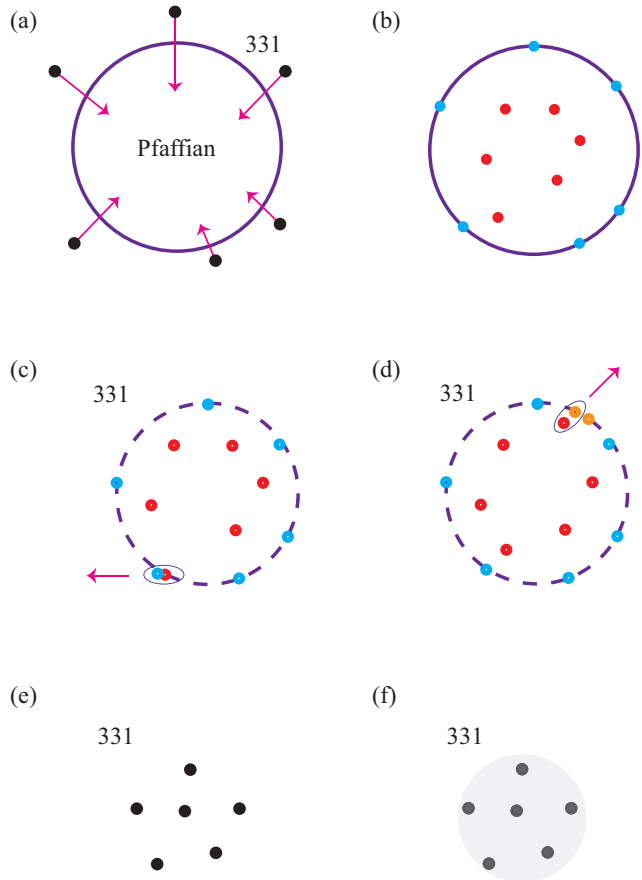}
\caption{Illustration of dragging quasiparticles in to the black hole (Pfaffian region) and simulating black hole evaporation in the 331-Pfaffian interface. Before the evaporation: (a) dragging Abelian charge-$e/4$ quasiparticles (black dots) in to the Pfaffian liquid; (b) creating neutral anyons $\tilde{\sigma}$ (blue dots) and non-Abelian charge-$e/4$ quasiparticles (red dots). Different mechanisms of releasing quasiparticles back to the 331 liquid during the evaporation: (c) combining a non-Abelian anyon with an existing $\tilde{\sigma}$ at the interface; (d) creating an additional pair of $\tilde{\sigma}$ (orange dots) and combining one of them with a non-Abelian anyon. After the evaporation: (e) the most idealistic scenario with the same number of quasiparticles as the initial configuration in (a); and (f) the generic situation with a superposition of different number of anyons in the Halperin-331 liquid. See the main text for more details.}
\label{fig:evaporation}
\end{figure}

In order to resolve the information paradox in our model, it is necessary to mimic a black hole evaporation in the 331-Pfaffian interface. As we will show, the process recovers the original information carried by the pseudospin degree of freedom naturally. To simplify the discussion, we assume only charge-$e/4$ Abelian quasiparticles were dragged across the interface before the ``evaporation". In general, one can also drag quasiholes and quasiparticles with other charges. Under the above assumption, we argued in Sec.~\ref{sec:matching} that neutral anyons $\tilde{\sigma}$ and non-Abelian charge-$e/4$ quasiparticles carrying $\sigma$ are created. This is illustrated in Figs.~\ref{fig:evaporation}(a) and~\ref{fig:evaporation} (b).

The ``black hole evaporation" is simulated by shrinking the Pfaffian region. Experimentally, it may be achieved by reducing the interlayer tunneling in the bilayer system~\cite{Suen1994, Papic2010, Sheng2016, Yang2017}. When a non-Abelian quasiparticle reaches the shrinking interface, it is released back to the Halperin-331 liquid. This process plays the role of Hawking radiation in the present setup. There are two different mechanisms for the conversion from a non-Abelian quasiparticle into an Abelian quasiparticle. First, the former may encounter an existing $\tilde{\sigma}$ at the interface. In this case, they recombine and transmute back to an Abelian quasiparticle [see Eqs.~\eqref{eq:inverted1} and~\eqref{eq:inverted3}]. This special scenario is shown in Fig.~\ref{fig:evaporation}(c). In order for this recombination to occur, it requires a highly delicate control of the shrinking process. Thus, it is unlikely to recombine all the existing $\tilde{\sigma}$ and $\sigma$ (with charge sector skipped) in this way. On the other hand, it is likely that a non-Abelian anyon reaches the interface at a position with no neutral anyon $\tilde{\sigma}$. Being already outside the Pfaffian liquid, the non-Abelian anyon still needs to transmute into an Abelian quasiparticle. In this case, a pair of $\tilde{\sigma}$ needs to be created at the interface. One of them combines with the non-Abelian anyon to covert into an Abelian quasiparticle in the Halperin-331 liquid. The remaining one is left at the interface. This mechanism is shown in Fig.~\ref{fig:evaporation}(d). Since the additional pair of $\tilde{\sigma}$ are created from the vacuum, they must have their fusion channel in the trivial topological sector. Hence, they do not carry additional information. All above processes are unitary, so information should be preserved.

\subsection{Recovery of Page curve and resolution of the information paradox}

Now, we show that the above ``black hole evaporation" recovers all original information and satisfies the Page curve. Here, the first subsystem consists of anyons remaining in the Pfaffian liquid and the interface. Another subsystem consists of Abelian quasiparticles in the Halperin-331 liquid. For simplicity, we call these two subsystems as (I) and (II), respectively. Since we treat the Pfaffian liquid as the black hole and the interface as an event horizon, reduced density matrices at different stages are obtained by partial tracing out (I). At the beginning of the evaporation, (II) is in a vacuum state with no quasiparticles, so its entropy is zero. When the Pfaffian liquid starts to shrink, the entropy of Abelian quasiparticles originating from their entanglement with (I) increases. However, the increase in entanglement entropy will not continue forever. By keep shrinking the Pfaffian region, the number of non-Abelian anyons and the dimension of the corresponding Hilbert space decrease. Hence, the dimensions of Hilbert spaces of (I) and (II) will become comparable and eventually equal to each other. The entanglement entropy reaches its maximum at this moment~\cite{Page1993-entropy}, which is known as the Page time. The Page time depends on the actual shrinking process. After passing the Page time, the entanglement entropy starts to decrease. 

In the most idealistic (yet most unlikely) situation which one can recombine all $N$ non-Abelian anyons with the originally existing $\tilde{\sigma}$ (exist before the evaporation) at the interface, the Page time occurs when $N/2$ quasiparticles are released to the 331 region. This feature does not hold in a generic situation. One can actually deduce the average entanglement entropy of (II) in the most idealistic case. We assume the initial state of the total system (before shrinking the Pfaffian region) is a random pure state $|\Psi\rangle$ in the $2^N$-dimensional Hilbert space. The entanglement entropy is averaged with respect to the unitary invariant Haar measure on the space of unitary vectors $|\Psi\rangle$ in the $2^N$-dimensional Hilbert space~\cite{Page1993-entropy}. Suppose $j$ non-Abelian anyons have been dragged out from the Pfaffian region and transmuted back to Abelian quasiparticles in the Halperin-331 liquid. The corresponding Hilbert space dimensions of (I) and (II) are given by $n=2^{N-j}$ and $m=2^j$, respectively. When $m\leq n$, the conjecture by Page (later proved by Sen~\cite{Sen1996}) suggests that the average entanglement entropy of (II) takes the form~\cite{Page1993-entropy},
\begin{eqnarray} \label{eq:Page-S}
\langle S_{\rm (II)}\rangle
\equiv S_{m,n}
=\left(\sum_{k=n+1}^{mn}\frac{1}{k}\right)-\frac{m-1}{2n}.
\end{eqnarray}
For $m>n$, one obtains $\langle S_{\rm (II)}\rangle$ by interchanging $m$ and $n$ in Eq.~\eqref{eq:Page-S}. By plotting $\langle S_{\rm (II)}\rangle$ versus $\ln{m}$, one concludes that it is identical to the one in Fig.~1 of Ref.~\cite{Page1993-BH}.

From our previous discussion, it is very likely that the number of $\tilde{\sigma}$ anyons increases during the ``black hole evaporation". This leads to two consequences. First, it is possible that all non-Abelian anyons in the Pfaffian liquid have been released to the 331 liquid, but some $\tilde{\sigma}$ anyons still remain at the interface. These anyons are entangled with the Abelian anyons in the Halperin-331 region, so the entanglement entropy is still nonzero. Second, the bound $N_L\sim L/\ell_B$ can be satisfied easily during the shrinking process. The discussion in Sec.~\ref{sec:bound} showed that pairs of Abelian quasiparticles and quasiholes with opposite charges will be released to the 331 region. After eliminating the Pfaffian region completely, the total number of anyons in the 331 region needs not be equal to the number in the original configuration (before sending the anyons in the Pfaffian liquid). Only in the most idealistic situation that we mentioned previously, these two numbers are equal as shown in Fig.~\ref{fig:evaporation}(a) and (e). In general, the final state of the system will have a superposition of different total numbers of anyons in the 331 liquid. It is reasonable since the total charge in the system is still conserved. This idea is illustrated in Fig.~\ref{fig:evaporation}(f). In particular, Fig.~\ref{fig:evaporation}(f) denotes a superposition state of $6+M$ charge-$e/4$ quasiparticles and $M$ charge-$-e/4$ quasiholes, where $M$ is a non-negative integer. This kind of superposition state is actually a closer analog of the actual Hawking radiation emitted from a black hole, which consists of different types of particles or excitations.

Independent of the actual shrinking process, the system must return to a pure state when the Pfaffian region is eliminated completely. Then, the entanglement entropy goes back to zero and resembles the Page curve. The original pseudospin information is recovered but in a highly entangled form. Thus, the paradox in our model is resolved. Our above discussion suggests that the Page curve in the present system should be more complicated than the one in Fig.~1 of Ref.~\cite{Page1993-BH}.

\section{Conclusion and discussion}

To conclude, we have identified and resolved an ``information paradox" in the 331-Pfaffian quantum Hall interface. The paradox originates from an apparent inability to recover the original pseudospin information of Abelian charge-$e/4$ quasiparticles after they cross the interface and enter the Pfaffian liquid. We employed the technique of anyon condensation and found that each incoming quasiparticle is transmuted into a pair of non-Abelian anyons. One of them is created in the Pfaffian liquid, whereas the other is created at the interface. Hence, the original information is stored nonlocally in the system, and cannot be recovered by any local measurement. We believe this is a fair analogy to an object falling into a real black hole, in the sense that while the information it carries is not lost, does become inaccessible to an (outside) observer. This resembles the idea of quantum information scrambling, which is consistent with the modern viewpoint that black holes are fast (perhaps the fastest) information scramblers~\cite{Hayden-Preskill, scrambler1, scrambler2, scrambler3, scrambler4}. The matching between the dimensions of Hilbert spaces for the Abelian quasiparticle and non-Abelian anyons further verifies the preservation of information.

Also, we considered the case when more quasiparticles are dragged across the interface. We argued that the maximum amount of information the system can store in a topologically protected way is bounded by the length of the interface. This feature is reminiscent of a similar bound in black hole set by its area due to the holographic principle and the Bekenstein entropy. Furthermore, we pointed out that the interface behaves like a firewall not only to anyons, but also to electrons. The latter is supported by observing an incoming electron from the Halperin-331 liquid can drip off its Fermi statistics at the interface. Finally, we discussed the simulation of black hole evaporation by shrinking the Pfaffian region which releases quasiparticles back to the 331 liquid. We argue explicitly that the corresponding entanglement entropy would follow the Page curve. As a result, the original pseudospin information is recovered and the ``information paradox" in our model is resolved. Note that remnants may be left at the end of evaporation in actual astrophysical black holes~\cite{remnant1, remnant2, remnant3}. This may provide an alternative resolution of the information paradox, which is not addressed in the present work.

It is quite surprising that the seemingly simple 331-Pfaffian interface has a rich analogy with black hole physics. At the same time, we need to point out some potential differences between our model and real astrophysical black holes. For a (semi)classical black hole, the horizon is not expected to have an effect on an infalling object (the so-called ``no drama scenario”), including the information carried by it. Whether this remains to be the case or not for a fully quantum-mechanical black hole is unclear. A firewall at the horizon is a possible scenario that is currently under investigation and debate~\cite{AMPS}. In our model, the Abelian quasiparticles must be transmuted when they cross the 331-Pfaffian interface. This is inevitable as the Halperin-331 and Pfaffian liquids allow different degrees of freedom. Thus, the interface in our model behaves like a firewall. In our opinion, this interface may be a very simple and accessible ``black hole firewall", which deserves more attention. In future work, it will be tempting to examine possible analogy of black hole thermodynamics in quantum Hall interfaces. It is also interesting to examine whether the 331-Pfaffian or other quantum Hall interfaces can provide an easy simulation of (a topological version of) the Hayden-Preskill protocol. 

The black-hole information paradox is arguably one of the most fundamental problems in physics, which involves gravitation, quantum field theory, and in particular, quantum information science. This long-standing problem is currently being actively studied by physicists in many different areas, and from very different perspectives (but so far only theoretically). Its resolution may well pave the way for the quantum theory of gravity, the holy grail of theoretical physics. While there is a lack of complete similarity between our model and certain “believed” processes in actual black holes (especially in the description of black hole evaporation which should be spontaneous), the analogy presented here provides a simple and accessible platform to simulate (i) apparent information loss, (ii) information scrambling, and (iii) information recovery. We believe these are arguably the most important and central concepts in understanding and resolving the original information paradox. Furthermore, our work may open a new research direction of studying how local information can be transmuted and stored nonlocally in an actual black hole. Since the concept of firewall and many other aspects in the paradox are still under intense debate, it is worthwhile to have simple analogies that capture some of the relevant concepts (but not necessarily all details precisely) in the original problem. In addition, our results have established a connection between quantum information, black hole physics and quantum Hall physics, and may bring experimentalists into this exciting research area.  

Lastly, it is worthwhile to mention that a deep connection between quantum Hall effect and gravitational physics has been revealed in previous work~\cite{Haldane2012, Son2016, Yang2016, Liou2019, Son2021, Gromov2021, Haldane2021, Stone2013, Vishveshwara2019, Vishveshwara2020}. In particular, Refs.~\cite{Stone2013, Vishveshwara2019, Vishveshwara2020} have suggested a possible simulation of Hawking-Unruh effect by scattering quasiparticles in quantum Hall systems. It is optimistic that more connections between black hole physics and quantum Hall physics may be discovered in the future.

\section*{Acknowledgment}

We thank N. E. Bonesteel for a useful discussion. This research was supported by the National Science Foundation Grant No. DMR-1932796, and performed at the National High Magnetic Field Laboratory, which is supported by National Science Foundation Cooperative Agreement No. DMR-1644779, and the State of Florida.


\begin{thebibliography}{99}

\bibitem{LIGO2016}
B. P. Abbott et al. (LIGO Scientific Collaboration and Virgo Collaboration),
\href{https://journals.aps.org/prl/abstract/10.1103/PhysRevLett.116.061102}
{Phys. Rev. Lett. \textbf{116}, 061102 (2016)}.

\bibitem{EHT2019}
The Event Horizon Telescope Collaboration et al,
\href{https://iopscience.iop.org/article/10.3847/2041-8213/ab0ec7}
{Astrophys. J. Lett. \textbf{875}, L1 (2019)}.

\bibitem{Paynter2021}
J. Paynter, R. Webster, and E. Thrane,
\href{https://www.nature.com/articles/s41550-021-01307-1}
{Nat. Astron. \textbf{5}, 560 (2021)}.

\bibitem{Hawking1974}
S. W. Hawking,
\href{https://www.nature.com/articles/248030a0}
{Nature \textbf{248}, 30 (1974)}.

\bibitem{Hawking1975}
S. W. Hawking,
\href{https://link.springer.com/article/10.1007%2FBF02345020}
{Commun. Math. Phys \textbf{43}, 199 (1975)}.

\bibitem{Hawking1976}
S. W. Hawking,
\href{https://journals.aps.org/prd/abstract/10.1103/PhysRevD.14.2460}
{Phys. Rev. D \textbf{14}, 2460 (1976)}.

\bibitem{soft-hair}
Notice that Hawking and his collaborators have recently introduced the idea of soft hair~\cite{HPS-hair}.

\bibitem{HPS-hair}
S. W. Hawking, M. J. Perry, and A. Strominger,
\href{https://journals.aps.org/prl/abstract/10.1103/PhysRevLett.116.231301}
{Phys. Rev. Lett. \textbf{116}, 231301 (2016)}.

\bibitem{Israel67}
W. Israel,
\href{https://journals.aps.org/pr/abstract/10.1103/PhysRev.164.1776}
{Phys. Rev. \textbf{164}, 1776 (1967)}.

\bibitem{Israel68}
W. Israel,
\href{https://link.springer.com/article/10.1007/BF01645859}
{Commun.Math. Phys. \textbf{8}, 245 (1968)}.

\bibitem{Carter71}
B. Carter,
\href{https://journals.aps.org/prl/abstract/10.1103/PhysRevLett.26.331}
{Phys. Rev. Lett. \textbf{26}, 331 (1971)}.

\bibitem{Giddings}
S. B. Giddings,
\href{https://arxiv.org/abs/hep-th/9508151}
{arXiv: hep-th/9508151}.

\bibitem{Mathur}
S. D. Mathur,
\href{https://iopscience.iop.org/article/10.1088/0264-9381/26/22/224001}
{Class. Quantum Gravity \textbf{26}, 224001 (2009)}.

\bibitem{Marolf}
D. Marolf,
\href{https://iopscience.iop.org/article/10.1088/1361-6633/aa77cc}
{Rep. Prog. Phys. \textbf{80}, 092001 (2017)}.

\bibitem{t-Hooft}
G. 't Hooft,
\href{https://arxiv.org/abs/gr-qc/9310026}
{arXiv:gr-qc/9310026}.

\bibitem{Susskind}
L. Susskind,
\href{https://aip.scitation.org/doi/10.1063/1.531249}
{J. Math. Phys. \textbf{36}, 6377 (1995)}.

\bibitem{Bousso}
R. Bousso,
\href{https://journals.aps.org/rmp/abstract/10.1103/RevModPhys.74.825}
{Rev. Mod. Phys. \textbf{74}, 825 (2002)}.

\bibitem{Barbon}
J. L. F. Barb\'{o}n,
\href{https://iopscience.iop.org/article/10.1088/1742-6596/171/1/012009}
{J. Phys.: Conf. Ser. \textbf{171}, 012009 (2009)}.

\bibitem{Maldacena}
J. M. Maldacena,
\href{https://link.springer.com/article/10.1023%2FA%3A1026654312961}
{Int. J. Theor. Phys. \textbf{38}, 1113 (1999)}.

\bibitem{AMPS}
A. Almheiri, D. Marolf, J. Polchinski, and J. Sully,
\href{https://link.springer.com/article/10.1007%2FJHEP02%282013%29062}
{J. High Energy Phys. \textbf{2013}, 62 (2013)}.

\bibitem{STU}
L. Susskind, L. Thorlacius and J. Uglum,
\href{https://journals.aps.org/prd/abstract/10.1103/PhysRevD.48.3743}
{Phys. Rev. D \textbf{48}, 3743 (1993)}.

\bibitem{StW}
C.R. Stephens, G. ’t Hooft and B.F. Whiting,
\href{https://iopscience.iop.org/article/10.1088/0264-9381/11/3/014}
{Class. Quantum Gravity \textbf{11}, 621 (1994)}.

\bibitem{CKW-inequality}
V. Coffman, J. Kundu and W. K. Wootters,
\href{https://journals.aps.org/pra/abstract/10.1103/PhysRevA.61.052306}
{Phys. Rev. A \textbf{61}, 052306 (2000)}.

\bibitem{Page1993-entropy}
D. N. Page,
\href{https://journals.aps.org/prl/abstract/10.1103/PhysRevLett.71.1291}
{Phys. Rev. Lett. \textbf{71}, 1291 (1993)}.

\bibitem{Page1993-BH}
D. N. Page,
\href{https://journals.aps.org/prl/abstract/10.1103/PhysRevLett.71.3743}
{Phys. Rev. Lett. \textbf{71}, 3743 (1993)}.

\bibitem{Page1993-review}
D. N. Page,
\href{https://arxiv.org/abs/hep-th/9305040}
{arXiv:hep-th/9305040}.

\bibitem{Hayden-Preskill}
P. Hayden and J. Preskill,
\href{https://iopscience.iop.org/article/10.1088/1126-6708/2007/09/120}
{J. High Energy Phys. \textbf{09}, 120 (2007)}.

\bibitem{Kitaev2017}
B. Yoshida and A. Kitaev,
\href{https://arxiv.org/abs/1710.03363}
{arXiv:1710.03363}.

\bibitem{Landsman2019}
K. A. Landsman, C. Figgatt, T. Schuster, N. M. Linke, B. Yoshida, N. Y. Yao, and C. Monroe,
\href{https://www.nature.com/articles/s41586-019-0952-6}
{Nature \textbf{567}, 61 (2019)}.

\bibitem{Yao2019}
B. Yoshida and N. Y. Yao,
\href{https://journals.aps.org/prx/abstract/10.1103/PhysRevX.9.011006}
{Phys. Rev. X \textbf{9}, 011006 (2019)}.

\bibitem{Penington}
G. Penington,
\href{https://arxiv.org/abs/1905.08255}
{arXiv:1905.08255}.

\bibitem{AEMM}
A. Almheiri, N. Engelhardt, D. Marolf, and H. Maxfield,
\href{https://link.springer.com/article/10.1007/JHEP12(2019)063}
{J. High Energy Phys. \textbf{2019}, 63 (2019)}.

\bibitem{PSSY}
G. Penington, S. H. Shenker, D. Stanford, and Z. Yang,
\href{https://arxiv.org/abs/1911.11977}
{arXiv:1911.11977}.

\bibitem{AEH}
C. Akers, N. Engelhardt, and D. Harlow,
\href{https://link.springer.com/article/10.1007%2FJHEP08%282020%29032}
{J. High Energy Phys. \textbf{2020}, 32 (2020)}.

\bibitem{AHMST}
A. Almheiri, T. Hartman, J. Maldacena, E. Shaghoulian, and A. Tajdini,
\href{https://link.springer.com/article/10.1007%2FJHEP05%282020%29013}
{J. High Energy Phys. \textbf{2020}, 13 (2020)}.

\bibitem{RMP2021}
A. Almheiri, T. Hartman, J. Maldacena, E. Shaghoulian, and A. Tajdini,
\href{https://journals.aps.org/rmp/abstract/10.1103/RevModPhys.93.035002}
{Rev. Mod. Phys. \textbf{93}, 35002 (2021)}.

\bibitem{Raju}
S. Raju,
\href{https://arxiv.org/abs/2012.05770}
{arXiv:2012:05770}.

\bibitem{Wen-book}
X. G. Wen, \textit{Quantum Field Theory of Many-Body Systems: From the Origin of Sound to an Origin of Light and Electrons} (Oxford University Press, Oxford, 2004).

\bibitem{Dima-review2020}
M. Heiblum and D. E. Feldman, \textit{Fractional Quantum Hall Effects: New Developments}, edited by B. I. Halperin and J. K. Jain (World Scientific, Singapore, 2020);
\href{https://www.worldscientific.com/doi/abs/10.1142/S0217751X20300094}
{Int. J. Mod. Phys. A \textbf{35}, 2030009 (2020)}.

\bibitem{Grosfeld2009}
E. Grosfeld and K. Schoutens,
\href{https://journals.aps.org/prl/abstract/10.1103/PhysRevLett.103.076803}
{Phys. Rev. Lett. \textbf{103}, 076803 (2009)}.

\bibitem{Bais-PRL2009}
F. A. Bais, J. K. Slingerland, and S. M. Haaker,
\href{https://journals.aps.org/prl/abstract/10.1103/PhysRevLett.102.220403}
{Phys. Rev. Lett. \textbf{102}, 220403 (2009)}.

\bibitem{wan16}
X. Wan and K. Yang,
\href{https://journals.aps.org/prb/abstract/10.1103/PhysRevB.93.201303}
{Phys. Rev. B {\bf 93}, 201303(R) (2016)}.

\bibitem{Yang2017}
K. Yang,
\href{https://journals.aps.org/prb/abstract/10.1103/PhysRevB.96.241305}
{Phys. Rev. B \textbf{96}, 241305(R) (2017)}.

\bibitem{Mross}
D. F. Mross, Y. Oreg, A. Stern, G. Margalit, and M. Heiblum,
\href{https://journals.aps.org/prl/abstract/10.1103/PhysRevLett.121.026801}
{Phys. Rev. Lett. {\bf 121}, 026801 (2018)}.

\bibitem{Wang}
C. Wang, A. Vishwanath, and B. I. Halperin,
\href{https://journals.aps.org/prb/abstract/10.1103/PhysRevB.98.045112}
{Phys. Rev. B {\bf 98}, 045112 (2018)}.

\bibitem{Lian}
B. Lian and J. Wang,
\href{https://journals.aps.org/prb/abstract/10.1103/PhysRevB.97.165124}
{Phys. Rev. B {\bf 97}, 165124 (2018)}.

\bibitem{simon20}
S. H. Simon, M. Ippoliti, M. P. Zaletel, and E. H. Rezayi,
\href{https://journals.aps.org/prb/abstract/10.1103/PhysRevB.101.041302}
{Phys. Rev. B {\bf 101}, 041302(R) (2020)}.

\bibitem{zhu20}
W. Zhu, D. N. Sheng, and K. Yang,
\href{https://journals.aps.org/prl/abstract/10.1103/PhysRevLett.125.146802}
{Phys. Rev. Lett. {\bf 125}, 146802 (2020)}.

\bibitem{Hughes2019}
J. May-Mann and T. L. Hughes,
\href{https://journals.aps.org/prb/abstract/10.1103/PhysRevB.99.155134}
{Phys. Rev. B \textbf{99}, 155134 (2019)}.

\bibitem{Regnault1}
V. Cr\'{e}pel, N. Claussen, N. Regnault, and B. Estienne,
\href{https://www.nature.com/articles/s41467-019-09169-y}
{Nat. Commun. \textbf{10}, 1860 (2019)}.

\bibitem{Regnault2}
V. Cr\'{e}pel, N. Claussen, B. Estienne, and N. Regnault,
\href{https://www.nature.com/articles/s41467-019-09168-z}
{Nat. Commun. \textbf{10}, 1861 (2019)}.

\bibitem{Nielsen1}
B. Jaworowski and A. E. B. Nielsen,
\href{https://journals.aps.org/prb/abstract/10.1103/PhysRevB.101.245164}
{Phys. Rev. B \textbf{101}, 245164 (2020)}.

\bibitem{Nielsen2}
B. Jaworowski and A. E. B. Nielsen,
\href{https://journals.aps.org/prb/abstract/10.1103/PhysRevB.103.205149}
{Phys. Rev. B \textbf{103}, 205149 (2021)}.

\bibitem{Teo2020}
R. Sohal, B. Han, L. H. Santos, and J. C. Y. Teo,
\href{https://journals.aps.org/prb/abstract/10.1103/PhysRevB.102.045102}
{Phys. Rev. B \textbf{102}, 045102 (2020)}.

\bibitem{Heiblum2021}
B. Dutta, W. Yang, R. A. Melcer, H. K. Kundu, M. Heiblum, V. Umansky, Y. Oreg, A. Stern, and D. Mross,
\href{https://www.science.org/doi/10.1126/science.abg6116}
{Science \textbf{375}, 193 (2021)}.

\bibitem{Mross2021}
M. Yutushui, A. Stern, and D. F. Mross,
\href{https://journals.aps.org/prl/abstract/10.1103/PhysRevLett.128.016401}
{Phys. Rev. Lett. \textbf{128}, 016401 (2022)}.

\bibitem{QH-interface2021}
Q. Li, K. K. W. Ma, R. Wang, Z.-X. Hu, H. Wang, and K. Yang,
\href{https://journals.aps.org/prb/abstract/10.1103/PhysRevB.104.125303}
{Phys. Rev. B \textbf{104}, 125303 (2021)}.

\bibitem{Halperin}
B. I. Halperin, Helv. Phys. Acta \textbf{56}, 75 (1983).

\bibitem{MR1991}
G. Moore and N. Read,
\href{https://www.sciencedirect.com/science/article/pii/055032139190407O}
{Nucl. Phys. B. \textbf{360}, 362 (1991)}.

\bibitem{TQC-RMP2008}
C. Nayak, S. H. Simon, A. Stern, M. Freedman, and S. Das Sarma,
\href{https://journals.aps.org/rmp/abstract/10.1103/RevModPhys.80.1083}
{Rev. Mod. Phys. \textbf{80}, 1083 (2008)}.

\bibitem{Suen-bilayer}
Y. W. Suen, L. W. Engel, M. B. Santos, M. Shayegan, and D. C. Tsui,
\href{https://journals.aps.org/prl/abstract/10.1103/PhysRevLett.68.1379}
{Phys. Rev. Lett. \textbf{68}, 1379 (1992)}.

\bibitem{Eisenstein-bilayer}
J. P. Eisenstein, G. S. Boebinger, L. N. Pfeiffer, K. W. West, and S. He,
\href{https://journals.aps.org/prl/abstract/10.1103/PhysRevLett.68.1383}
{Phys. Rev. Lett. \textbf{68}, 1383 (1992)}.

\bibitem{Papic2010}
Z. Papi\'{c}, M. O. Goerbig, N. Regnault, and M. V. Milovanovi\'{c},
\href{https://journals.aps.org/prb/abstract/10.1103/PhysRevB.82.075302}
{Phys. Rev. B \textbf{82}, 075302 (2010)}.

\bibitem{Sheng2016}
W. Zhu, Zhao Liu, F. D. M. Haldane, and D. N. Sheng,
\href{https://journals.aps.org/prb/abstract/10.1103/PhysRevB.94.245147}
{Phys. Rev. B \textbf{94}, 245147 (2016)}.

\bibitem{Suen1994}
Y. W. Suen, H. C. Manoharan, X. Ying, M. B. Santos, and M. Shayegan,
\href{https://journals.aps.org/prl/abstract/10.1103/PhysRevLett.72.3405}
{Phys. Rev. Lett. \textbf{72}, 3405 (1994)}.

\bibitem{Yang-book}
For a general introduction to quantum Hall effect, readers may refer to
S. M. Girvin and K. Yang, \textit{Modern Condensed Matter Physics} (Cambridge University Press, Cambridge, 2019). In particular, Chaps. 12 and 16.

\bibitem{review-fractional}
D. E. Feldman and B. I. Halperin,
\href{https://iopscience.iop.org/article/10.1088/1361-6633/ac03aa}
{Rep. Prog. Phys. \textbf{84}, 076501 (2021)}.

\bibitem{Tsui1982}
D. C. Tsui, H. L. Stormer, and A. C. Gossard,
\href{https://journals.aps.org/prl/abstract/10.1103/PhysRevLett.48.1559}
{Phys. Rev. Lett. \textbf{48}, 1559 (1982)}.

\bibitem{Laughlin}
R. B. Laughlin,
\href{https://journals.aps.org/prl/abstract/10.1103/PhysRevLett.50.1395}
{Phys. Rev. Lett. \textbf{50}, 1395 (1983)}.

\bibitem{Arovas}
Daniel Arovas, J. R. Schrieffer, and Frank Wilczek,
\href{https://journals.aps.org/prl/abstract/10.1103/PhysRevLett.53.722}
{Phys. Rev. Lett. \textbf{53}, 722 (1984)}.

\bibitem{de-Picciotto}
R. de Picciotto, M. Reznikov, M. Heiblum, V. Umansky, G. Bunin, and D. Mahalu,
\href{https://www.nature.com/articles/38241}
{Nature \textbf{389}, 162, (1997)}.

\bibitem{Saminadayar}
L. Saminadayar, D. C. Glattli, Y. Jin, and B. Etienne,
\href{https://journals.aps.org/prl/abstract/10.1103/PhysRevLett.79.2526}
{Phys. Rev. Lett. \textbf{79}, 2526 (1997)}.

\bibitem{Bartolomei}
H. Bartolomei, M. Kumar, R. Bisognin, A. Marguerite, J.-M. Berroir, E. Bocquillon, B. Pla\c{c}ais, A. Cavanna, Q. Dong, U. Gennser, and Y. Jin,
\href{https://science.sciencemag.org/content/368/6487/173}
{Science \textbf{368}, 6487 (2020)}.

\bibitem{CFT-QHE}
T. H. Hansson, M. Hermanns, S. H. Simon, and S. F. Viefers,
\href{https://journals.aps.org/rmp/abstract/10.1103/RevModPhys.89.025005}
{Rev. Mod. Phys. \textbf{89}, 025005 (2017)}.

\bibitem{MR-edge}
M. Milovanovi\'{c} and N. Read,
\href{https://journals.aps.org/prb/abstract/10.1103/PhysRevB.53.13559}
{Phys. Rev. B \textbf{53}, 13559 (1996)}.

\bibitem{Levin-APf}
M. Levin, B. I. Halperin, and B. Rosenow,
\href{https://journals.aps.org/prl/abstract/10.1103/PhysRevLett.99.236806}
{Phys. Rev. Lett. \textbf{99}, 236806 (2007)}.

\bibitem{Lee-APf}
S.-S. Lee, S. Ryu, C. Nayak, and M. P. A. Fisher,
\href{https://journals.aps.org/prl/abstract/10.1103/PhysRevLett.99.236807}
{Phys. Rev. Lett. \textbf{99}, 236807 (2007)}.

\bibitem{ZF2016}
P. T. Zucker and D. E. Feldman,
\href{https://journals.aps.org/prl/abstract/10.1103/PhysRevLett.117.096802}
{Phys. Rev. Lett. \textbf{117}, 096802 (2016)}.

\bibitem{Read-Green}
N. Read and D. Green,
\href{https://journals.aps.org/prb/abstract/10.1103/PhysRevB.61.10267}
{Phys. Rev. B \textbf{61}, 10267 (2000)}.

\bibitem{Nayak-Wilczek}
C. Nayak and F. Wilzcek,
\href{https://www.sciencedirect.com/science/article/pii/0550321396004300}
{Nucl. Phy. B \textbf{479}, 529 (1996)}.

\bibitem{CFT-book}
P. Di Francesco, P. Mathieu, and D. Senechal, \textit{Conformal Field Theory} (Springer-Verlag, New York, 1997).

\bibitem{Bais-PRB2009}
F. A. Bais and J. K. Slingerland,
\href{https://journals.aps.org/prb/abstract/10.1103/PhysRevB.79.045316}
{Phys. Rev. B \textbf{79}, 045316 (2009)}.

\bibitem{Ellens2014}
I. S. Eli\"{e}ns,  J. C. Romers, and F. A. Bais,
\href{https://journals.aps.org/prb/abstract/10.1103/PhysRevB.90.195130}
{Phys. Rev. B \textbf{90}, 195130 (2014)}.

\bibitem{Burnell-review}
F. J. Burnell,
\href{https://www.annualreviews.org/doi/10.1146/annurev-conmatphys-033117-054154}
{Annu. Rev. Condens. Matter Phys. \textbf{9}, 307 (2017)}.

\bibitem{Bernevig}
T. Neupert, H. He, C. von Keyserlingk, G. Sierra, and B. A. Bernevig,
\href{https://journals.aps.org/prb/abstract/10.1103/PhysRevB.93.115103}
{Phys. Rev. B \textbf{93}, 115103 (2016)}.

\bibitem{footnote-gap}
The gapping of modes is realized when the tunneling amplitude in $H_T$ is a constant instead of a random variable in Eq.~\eqref{eq:mode-gap}. Strictly speaking, this kind of nonrandom tunneling is unphysical as there is a momentum mismatch between different modes. However, gapped modes and localized modes discussed in Sec.~\ref{sec:331-Pf} have the same physical consequences in the present discussion.

\bibitem{footnote0}
In our opinion, this is a big advantage over the Pfaffian-NASS interface~\cite{Bais-PRL2009, Grosfeld2009}. In that case, the NASS quantum Hall liquid at $\nu=4/7$ has quasiparticle with the smallest charge of $e/7$. Hence, a net charge of $e/4-e/7=3e/28$ is absorbed, which leads to an accumulation of charges at the interface. Furthermore, the difference between filling factors of the two quantum Hall liquids may complicate the gapping of counterpropagating charge modes at the interface. This difference also introduces additional challenges in realizing the interface experimentally.

\bibitem{Ardonne-spin}
T. M\r{a}nsson, V. Lahtinen, J. Suorsa, and E. Ardonne,
\href{https://journals.aps.org/prb/abstract/10.1103/PhysRevB.88.041403}
{Phys. Rev. B \textbf{88}, 041403(R) (2013)}.

\bibitem{Ardonne-para}
V. Lahtinen, T. M\r{a}nsson, and E. Ardonne,
\href{https://scipost.org/SciPostPhysCore.4.2.014}
{SciPost Phys. Core \textbf{4}, 014 (2021)}.

\bibitem{DVVV1989}
R. Dijkgraaf, C. Vafa, E. Verlinde, H. Verlinde,
\href{https://link.springer.com/article/10.1007/BF01238812}
{Comm. Math. Phys. \textbf{123}, 485 (1989)}.

\bibitem{footnote}
Although Fig.~\ref{fig:interface} may imply the interface is a one-dimensional object, it is not. The interface actually penetrates in both quantum Hall liquids and has a finite width (but much shorter than its length), similar to the interface between Pfaffian and anti-Pfaffian liquids~\cite{zhu20}. Therefore, braiding of anyons can occur at the interface.

\bibitem{Chamon2019}
Z.-C. Yang, K. Meichanetzidis, S. Kourtis, and C. Chamon,
\href{https://journals.aps.org/prb/abstract/10.1103/PhysRevB.99.045132}
{Phys. Rev. B \textbf{99}, 045132 (2019)}.

\bibitem{footnote2}
In the Pfaffian-NASS interface, one may consider a similar ``information paradox" by dragging a charge-$e/7$ non-Abelian quasiparticle from the NASS liquid to the Pfaffian liquid. Depending on the species, the neutral sector is given by $\sigma_{\uparrow}$ or $\sigma_{\downarrow}$ in the $c=6/5$ CFT. Both transmute into $(\bar{\sigma}, \sigma)$, where $\bar{\sigma}$ is a primary field in the $\mathcal{M}(5,4)$ minimal model describing the interface~\cite{Bais-PRL2009, Grosfeld2009}. Since $\bar{\sigma}$ has a quantum dimension $d=(1+\sqrt{5})/\sqrt{2}>\sqrt{2}$, the Hilbert space of anyons at the interface has a larger dimension.

\bibitem{Preskill-notes}
 J. Preskill, Lecture Notes for Physics 219: Quantum Computation – Chapter 9. Topological Quantum Computation.

\bibitem{TQC}
M. H. Freedman, A. Kitaev, M. J. Larsen, and Z. Wang,
\href{https://www.ams.org/journals/bull/2003-40-01/S0273-0979-02-00964-3/home.html}
{Bull. Amer. Math. Soc. \textbf{40}, 31 (2003)}.

\bibitem{Kitaev2003}
A. Y. Kitaev,
\href{https://www.sciencedirect.com/science/article/pii/S0003491602000180}
{Ann. Phys. \textbf{303}, 2 (2003)}.

\bibitem{foot-TQC}
It is reminded that one cannot perform universal TQC (or reproduce any unitary operation to arbitrary accuracy by braiding of anyons) by using anyons in the Ising CFT~\cite{Werner2009, Pachos}. More complicated anyons such as Fibonacci anyons are required for achieving the universality~\cite{Bonesteel}.

\bibitem{Bekenstein72}
J. D, Bekenstein,
\href{https://link.springer.com/article/10.1007%2FBF02757029}
{Lett. Nuovo Cimento \textbf{4}, 737 (1972)}.

\bibitem{Bekenstein73}
J. D. Bekenstein,
\href{https://journals.aps.org/prd/abstract/10.1103/PhysRevD.7.2333}
{Phys. Rev. D \textbf{7}, 2333 (1973)}.

\bibitem{Bekenstein74}
J. D. Bekenstein,
\href{https://journals.aps.org/prd/abstract/10.1103/PhysRevD.9.3292}
{Phys. Rev. D \textbf{9}, 3292 (1974)}.

\bibitem{BBK2014}
M. Barkeshli, E. Berg, and S. Kivelson,
\href{https://science.sciencemag.org/content/346/6210/722}
{Science \textbf{346}, 722 (2014)}.

\bibitem{Sen1996}
S. Sen,
\href{https://journals.aps.org/prl/abstract/10.1103/PhysRevLett.77.1}
{Phys. Rev. Lett. \textbf{77}, 1 (1996)}.

\bibitem{scrambler1}
Y. Sekino and L. Susskind,
\href{https://iopscience.iop.org/article/10.1088/1126-6708/2008/10/065}
{J. High Energy Phys. \textbf{2008}, 65 (2008)}.

\bibitem{scrambler2}
L. Susskind,
\href{https://arxiv.org/abs/1101.6048}
{arXiv:1101.6048}.

\bibitem{scrambler3}
J. L. F. Barb\'{o}n and J. M. Mag\'{a}n,
\href{https://journals.aps.org/prd/abstract/10.1103/PhysRevD.84.106012}
{Phys. Rev. D \textbf{84}, 106012 (2011)}.

\bibitem{scrambler4}
N. Lashkari, D. Stanford, M. Hastings, T. Osborne and P. Hayden, 
\href{https://link.springer.com/article/10.1007%2FJHEP04%282013%29022}
{J. High Energy Phys. \textbf{2013}, 22 (2013)}.

\bibitem{remnant1}
R. J. Adler, P. Chen, D. I. Santiago, 
\href{https://link.springer.com/article/10.1023/A:1015281430411}
{Gen. Rel. Grav. \textbf{33}, 2101 (2001)}.

\bibitem{remnant2}
Y. Aharonov, A. Casher, and S. Nussinov, 
\href{https://www.sciencedirect.com/science/article/pii/0370269387913207}
{Phys. Lett. B \textbf{191}, 51 (1987)}.

\bibitem{remnant3}
P. Chen, Y. C. Ong, D.-h. Yeom,
\href{https://www.sciencedirect.com/science/article/pii/S0370157315004391}
{Phys. Rep. \textbf{603}, 1 (2015)}.

\bibitem{Haldane2012}
B. Yang, Z.-X. Hu, Z. Papi\'{c}, and F. D. M. Haldane,
\href{https://journals.aps.org/prl/abstract/10.1103/PhysRevLett.108.256807}
{Phys. Rev. Lett. \textbf{108}, 256807 (2012)}.

\bibitem{Son2016}
S. Golkar, D. X. Nguyen and D. T. Son,
\href{https://link.springer.com/article/10.1007/JHEP01(2016)021}
{JHEP \textbf{01}, 21 (2016)}.

\bibitem{Yang2016}
K. Yang,
\href{https://journals.aps.org/prb/abstract/10.1103/PhysRevB.93.161302}
{Phys. Rev. B \textbf{93}, 161302(R) (2016)}.

\bibitem{Liou2019}
S.-F. Liou, F. D. M. Haldane, K. Yang, and E. H. Rezayi,
\href{https://journals.aps.org/prl/abstract/10.1103/PhysRevLett.123.146801}
{Phys. Rev. Lett. \textbf{123}, 146801 (2019)}.

\bibitem{Son2021}
D. X. Nguyen and D. T. Son,
\href{https://journals.aps.org/prresearch/abstract/10.1103/PhysRevResearch.3.023040}
{Phys. Rev. Research \textbf{3}, 023040 (2021)}.

\bibitem{Gromov2021}
Z. Liu, A. C. Balram, Z. Papi\'{c}, and A. Gromov,
\href{https://journals.aps.org/prl/abstract/10.1103/PhysRevLett.126.076604}
{Phys. Rev. Lett. \textbf{126}, 076604 (2021)}.

\bibitem{Haldane2021}
F. D. M. Haldane, E. H. Rezayi, and K. Yang,
\href{https://journals.aps.org/prb/abstract/10.1103/PhysRevB.104.L121106}
{Phys. Rev. B \textbf{104}, L121106 (2021)}.

\bibitem{Stone2013}
M. Stone,
\href{https://iopscience.iop.org/article/10.1088/0264-9381/30/8/085003}
{Class. Quantum Gravity \textbf{30}, 085003 (2013)}.

\bibitem{Vishveshwara2019}
S. S. Hegde, V. Subramanyan, B. Bradlyn, and S. Vishveshwara,
\href{https://journals.aps.org/prl/abstract/10.1103/PhysRevLett.123.156802}
{Phys. Rev. Lett. \textbf{123}, 156802 (2019)}.

\bibitem{Vishveshwara2020}
V. Subramanyan, S. S. Hegde, S. Vishveshwara, and B. Bradlyn,
\href{https://arxiv.org/abs/2012.09875}
{arXiv:2012.09875}.

\bibitem{Werner2009}
A. Ahlbrecht, L. S. Georgiev, and R. F. Werner,
\href{https://journals.aps.org/pra/abstract/10.1103/PhysRevA.79.032311}
{Phys. Rev. A \textbf{79}, 032311 (2009)}.

\bibitem{Pachos}
V. Lahtinen and J. K. Pachos,
\href{https://scipost.org/10.21468/SciPostPhys.3.3.021}
{SciPost Phys. \textbf{3}, 021 (2017)}.

\bibitem{Bonesteel}
N. E. Bonesteel, L. Hormozi, G. Zikos, and S. H. Simon,
\href{https://journals.aps.org/prl/abstract/10.1103/PhysRevLett.95.140503}
{Phys. Rev. Lett. \textbf{95}, 140503 (2005)}.

\end{thebibliography}
\end{document}